# Elucidating the Role of Prelithiation in Si-based Anodes for Interface Stabilization

Shuang Bai,[1] Wurigumula Bao,[2,3] Kun Qian,[2] Bing Han,[2] Weikang Li,[2] Baharak Sayahpour,[1] Bhagath Screenarayanan,[2] Darren H.S. Tan,[2] So-yeon Ham,[1] and Ying Shirley Meng[1,2,3,*]

1. Materials Science and Engineering, University of California San Diego, La Jolla, CA 92093, United States.
2. Department of NanoEngineering, University of California San Diego, La Jolla, CA 92093, United States.
3. Pritzker School of Molecular Engineering, University of Chicago, Chicago, IL 60637, United States.

Corresponding author: shirleymeng@uchicago.edu

# Abstract

Prelithiation as a facile and effective method to compensate the lithium inventory loss in the initial cycle has progressed considerably both on anode and cathode sides. However, much less research has been devoted to the prelithiation effect on the interface stabilization for long-term cycling of Si-based anodes. An in-depth quantitative analysis of the interface that form during the prelithiation of $SiO_x$ is presented here and the results are compared with prelithiaton of Si anodes. Local structure probe combined with detailed electrochemical analysis reveals that a characteristic mosaic interface is formed on both prelithiated $SiO_x$ and Si anodes. This mosaic interface containing multiple lithium silicates phases, is fundamentally different from the solid electrolyte interface (SEI) formed without prelithiation. The ideal conductivity and mechanical properties of lithium silicates enable improved cycling stability of both prelithiated anodes. With a higher ratio of lithium silicates due to the oxygen participation, prelithiated $SiO_{1.3}$ anode improves the initial coulombic efficiency to 94% in full cell and delivers good cycling retention after hundreds cycles under lean electrolyte conditions. The insights provided in this work could be used to further optimize high Si loading based anode in future high energy density batteries.

# INTRODUCTION

The past several decades has witnessed conventional lithium-ion batteries (LIBs) dominate the portable devices and consumer electronics market. Today, LIBs are rapidly penetrating other technologies including electric vehicles and grid storage. The success of LIBs stems from the rapidly growing efforts in battery research and development, leading to vast improvements in materials performance and decrease in production costs. The justification for a more widespread adoption of LIBs entails overcoming fundamental obstacles such as demands for higher energy and power density, cycle life, less safety hazards and lowered costs of batteries per kWh. For instance, LIBs need to achieve energy densities of more than 350 Wh/kg to better meet market demands.[1] Considering the fact that most inactive components of LIBs (such as current collector, separator, packaging, etc.) have already been fully optimized with decades of industrial commercialization and large-scale production of LIBs, the main performance improvements now lie within improving the electrode and electrolyte materials for energy density, lifetime, safety and costs.

Compared to the cathode development, which has seen extensive improvements in both reversible capacity and nominal cell voltage,[2] the anode has not experienced as much progress from traditional graphite-based anodes used in today's LIBs. Si-based anodes exhibit high theoretical specific capacities (3580 mAh/g for Si and >2000 mAh/g for $SiO_x$ depending on x value) and potential close to Li metal (0.3 V vs $Li/Li^+$),[3,4] making it capable of realizing high energy density in LIBs. Moreover, its relative elemental abundance and environmental friendliness make it both cost effective and sustainable for large scale adoption in LIBs. Unlike Li metal, it is also stable under ambient conditions and can be processed without use of inert gas environments, further lowering potential costs of manufacturing.

Despite the high theoretical energy density of Si (**Figure S1**), the poor cycling performance resulting from the continuous interfacial growth and Li inventory consumption leads to poor active material utilization in the commercial level applications.[5] Together with the high cost of mass production, the commercialization of nanosized silicon as an anode material for lithium-ion batteries has been largely impeded. $SiO_x$ emerging as an alternative affordable, high energy density anode draws great interest for the good cycling stability and active material utilization. However, the initial coulombic efficiency (ICE) of silicon-based anodes is relatively low, which is associated with the electrolyte decomposition, formation of solid electrolyte interface (SEI), irreversible parasitic reactions (such as $Li_2O$ and lithium silicates formation for $SiO_x$);[6]

Even though strategies like innovated electrolytes have been reported, lithium loss in the initial cycle of Si-based materials is way higher than that of the commercial graphite.[7–9] This higher active lithium loss is detrimental to the specific energy density of the full cell consisting of a silicon-based anode and a positive electrode with limited lithium amount. Therefore, a promising approach, prelithiation, by presetting lithium ions in the anode material before electrochemical cycling is effective in compensating for the initial Li loss, resulting in improved energy density of full cells.[10] Based on the different protocols and mechanisms, the prelithiation strategies on the Si-based anodes can be divided into two categories: electrochemical method and chemical method.[11] The chemical method, including mechanical alloy and one-pot metallurgy by mixing anode materials (e.g., Si,[12] $SiO_x$[13]) and Li source, is scalable to prepare lithiated anode materials. However, it is difficult to accurately control the degree of prelithiation with the chemical method and prepare electrodes on a large scale due to the sensitivity of the lithiated anode materials to the environment. The electrochemical method by architecting half cell structure composed of Si-based electrodes and Li metal foil is widely used in lab scale, in which the prelithiation amount can be controlled by the voltage or the prelithiation time.[14,15] To avoid complex operations caused by half cell electrochemical method, an advanced strategy was proposed to contact the Si-based anode directly with Li metal foil and electrolyte in between.[16] The direct contact method creates short-circuit, and the electrons will move immediately at the point of contact under the action of the electric field. In order to remain electrically neutral, the Li metal will release lithium ions through the electrolyte into the anode material to complete the lithiation process. This short-circuit electrochemical method utilizing the self-discharge of lithium metal can be processed in a simple assembly and the prelithiation time can reflect the pre-lithiation degree mediately.

Although there is no doubt that the prelithiation of Si-based anodes can improve the ICE, as shown in **Table S1**,[16,17,26,18–25] the impact of prelithiation on long-term cycling stability in the full cell remains elusive. Kim et al. reported the capacity retention of the prelithiated $SiO_x$/NCA ($Li[Ni_{0.8}Co_{0.15}Al_{0.05}]O_2$) full cell was 15% lower than the pristine counterpart after 100 cycles.[16] By using the prelithiated $SiO_x$, Chung et al. enhanced the energy density of full cell by 50% compared to that adopting pristine $SiO_x$ with similar cycling retention (70% for the prelithiated anode versus 75% for the pristine anode) over 800 cycles.[21] While cycling stability of full cells using prelithiated anodes were improved in other works,[17–20] the improvement was generally ascribed to the stable SEI layer established during the prelithiation process. Shen et al. applied X-ray photoelectron spectroscopy (XPS) to pinpoint the prelithiation products and it has been found that the SEI mainly contains $Li_2CO_3$ and LiF,[27] which has little difference compared to the components of SEI formed in the electrochemical reaction. The controversial results on the cycling stability by using the prelithiated Si-based anodes reflects the significant complexity of prelithiation process and challenges in characterization of the prelithiated products: 1) Products are of low crystallinity; 2) Multiple components are present, which is expected to depend on the degree of prelithiation. Even the SEI composition and structure formed during the prelithiation process remain elusive, these are fundamental aspects that underpin the mechanism(s) of the interface stabilization.

In this work, we systematically investigate the SEI components, structure, and properties for the prelithiation process in two representative Si-based anode materials, micro silicon (μSi) and silicon monoxide ($SiO_x$), to reveal insights into the prelithiation effects on the cycling stability of full cells with $LiFePO_4$ (LFP) cathode materials. Micron sized Si and $SiO_x$ are selected in this study due to their cost effectiveness. The short-circuit electrochemical method is applied in this study with high pressure (~ 5.5 kPa) onto Li metal foil and Si-based anodes to enable sufficient contact. After electrochemical analysis and X-ray diffraction (XRD) to characterize the prelithiated materials, we apply *ex situ* transmission electron microscopy (TEM) under cryogenic temperature to identify crystalline SEI components and their spatial distribution. In addition, XPS depth profiling is used to semi-quantitatively analyze the SEI composition for both crystalline and amorphous phases. We then compare the SEI composition and structure obtained from the prelithiation and electrochemical process, so as to investigate the physical properties of SEI and their impact on the interface stabilization over the cycling.

## RESULTS & DISCUSSION

### ICE Improvement of Si-based Anode Materials by Prelithiation

To realize high volumetric energy density, micro sized Si and $SiO_x$ are applied in the anode fabrication with high active material ratios (70% by weight). Unlike the diamond cubic crystal structure of μSi, the powder XRD pattern of $SiO_x$ (**Figure S2**) contains no sharp Bragg reflections and only diffuse scattering from the amorphous components are visible. To investigate the microstructure of pristine $SiO_x$, electron energy loss spectroscopy (EELS) was used and the results are shown in **Figure S3**. The EELS spectra of the Si L-edge from the surface and bulk region of the particle, as indicated in high angle annular dark field (HAADF) images in **Figure S3**, is compared with the spectra acquired from the standard samples of Si and $SiO_2$. The comparison clearly indicates $SiO_x$ contains two separate phases, Si and $SiO_2$. The Si local environment is similar between the bulk and surface region. These results are consistent with the proposed microstructure model, in which $SiO_x$ consists of Si and $SiO_2$ including the sub-oxide interphase boundary layer making up 20-25% of the total number of atoms.[28]

The oxygen content in $SiO_x$ has a critical impact on structural changes that occur during electrochemical cycling.[29] The atomic ratio between O and Si is determined to be 1.32(2) by inductively coupled plasma mass spectrometry (ICP-MS) following the similar protocol as reported.[30] Prelithiated Si and $SiO_{1.3}$ anodes were then prepared by the short-circuit electrochemical method using the setup provided in **Figure S4**. Note external pressure is necessary to deal with the nonuniform prelithiation issues due to the uneven contact between silicon-based anodes and lithium metal. Pristine and prelithiated $SiO_{1.3}$ and μSi electrodes with different prelithiation time were then galvanostatically discharged to 50 mV and then charged to 1.5 V in half cell with Li metal as the counter electrode. As shown in **Figure 1A and 1B**, the charge (delithiation) capacity of prelithiated electrodes is similar to the pristine electrode while the discharge (lithiation) capacity of prelithiated electrodes is lower than the pristine electrode. This indicates the insertion of lithium during the prelithiation process compensates the lithium loss derived from the formation of irreversible byproducts, but hardly affect the reversible capacity. The optimized prelithiation time is 12.5 mins for $SiO_{1.3}$ when the ICE reaches 96%, while shorter time (4 mins) is required for Si to achieve the optimized ICE (101%). This difference is due to the thinner mass loading of Si anode to keep the same negative to positive electrode capacity ratio (N/P ratio) of 1.1-1.2 in the full cell setup.

The differential capacity curves for the non-prelithiated $SiO_{1.3}$ in **Figure 1C** display three distinctive peaks at above 0.3 V (vs. Li/Li$^+$), 0.24 V and ~0.1 V upon discharging. The irreversible peaks above 0.3 V are related to the electrolyte decomposition process with the formation of byproducts, such as $Li_2CO_3$ and LiF. The second peak could correspond to the formation of lithium silicates and lithium oxide.[31] Upon further discharge of $SiO_{1.3}$ to ~0.1 V, the redox reaction could be assigned to Li-Si alloy phase formation. This redox potential is similar to that seen for μSi when the discharge voltage reaches 0.1 V, suggesting similar $Li_xSi$ reactions occur. Only the redox reactions at 0.24 V and 0.1 V are partially reversible during the charge process with potential shifts to 300 and 450 mV, respectively. The irreversible reaction peaks disappear and the peaks below 0.3 V decrease for the prelithiated $SiO_{1.3}$, which indicates the prelithiation products contain both irreversible and reversible phases. The XRD pattern for the prelithiated $SiO_{1.3}$ (**Figure S2**) contains Bragg reflections from $Li_{13}Si_4$, $Li_4SiO_4$, $Li_2SiO_3$, and $Li_2O$ phases, which is consistent with the differential capacity analysis. The electrochemical reactions of the μSi electrode are characterized by redox reactions observed at 0.1 V for the discharge and at 0.45 V in the subsequent charge process (**Figure 1D**). The presence of these clear voltage processes indicates that these reactions essentially proceed via two-phase reactions, whereas the reversible reactions correspond to the conversion between the crystalline Si structure and the amorphous $Li_xSi$ or crystalline $Li_{15}Si_4$ phase. The prelithiation process of µSi mainly reduces the capacity from the Li-Si alloy reaction during the discharge.

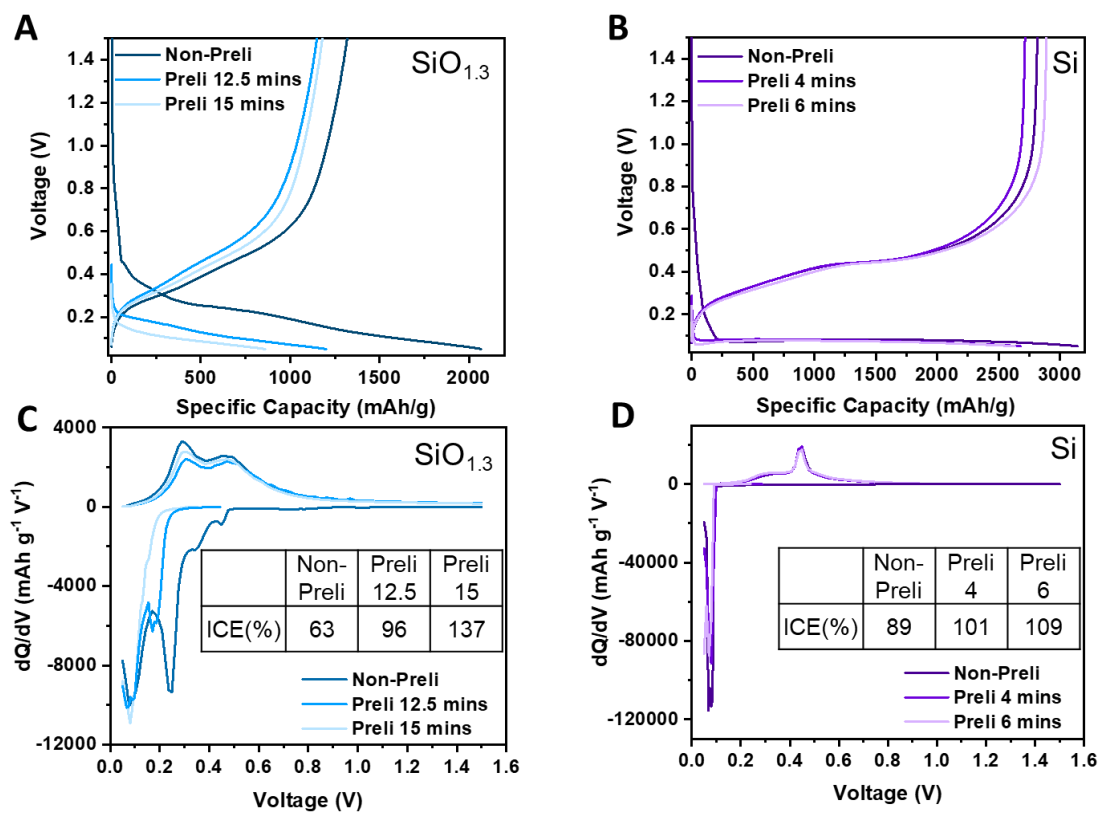

**Figure 1.** Voltage profile of $SiO_{1.3}$ (A) for the first cycle and (C) corresponding dQ/dV analysis in half cell with and without prelithiation. Voltage profile of µSi (B) for the first cycle and (D) corresponding dQ/dV analysis in half cell with and without prelithiation.

The improvement of ICE from this work is then compared with other published results, as shown in **Figure 2**. For a fair comparison, all data are from half cells tested with a similar voltage range and C-rate. Data are available in **Table S1**. Our work not only demonstrates the highest active material ratio for both prelithiated µSi and $SiO_x$ anodes, but also delivers the ideal ICE (close to 100%) which is critical for the full cell cycling stability evaluation. The inconsistent full cell cycling stability improvement by using prelithiated Si-based anodes in the published results may be ascribed to the lower ICE (**Figure 2**) and indelicate control of the prelithiation degree.

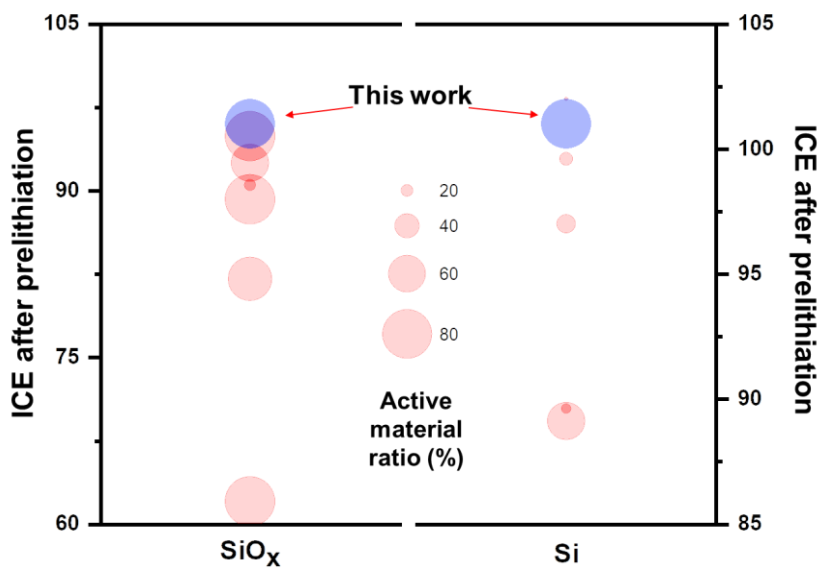

**Figure 2.** Literature summary on the ICE in half cell of prelithiated $SiO_x$ and Si.[16,17,26,18–25]

**Improvement of Full Cell Cycling Stability by Prelithiated Si-based Anodes**

The prelithiated $SiO_{1.3}$ and μSi were then paired with the LFP cathode to study the prelithiation effect on the long-term cycling stability of full cells. Olivine-type LFP was picked as the cathode active material in the full cell testing due to the structural stability. As shown in **Figure S5**, the LFP cathode with active material loading ~ 1 mAh/cm² can deliver 91% capacity retention after 200 cycles in half cell. The full cell cycling performance degradation can thus be reasonably correlated with the changes from Si-based anode side. The full cells using prelithiated $SiO_{1.3}$ and μSi deliver almost the same ICE of 94% (**Figure 3A and 3B**). This ICE improvement compared with the full cell using pristine anodes (44% for $SiO_{1.3}$ and 79% for μSi) is expected due to the delicate control of the prelithiation process. More importantly, the capacity retention of the full cell improves to 77% from 55% after 200 cycles by using prelithiated $SiO_{1.3}$ with 0.7% CE increase for every cycle on average (**Figure 3C and 3E**). The same improvement trend is observed for the full cell using prelithiated μSi (**Figure 3D and 3E**), where the capacity retention increases from 24% to 44%. Despite the improvement, the full cell using prelithiated Si shows much lower capacity retention (44% versus 77%) and average CE (99.5% versus 99.9%) compared with that using prelithiated $SiO_{1.3}$. This obvious difference can be explained from both the bulk structure evolution and interface stability.

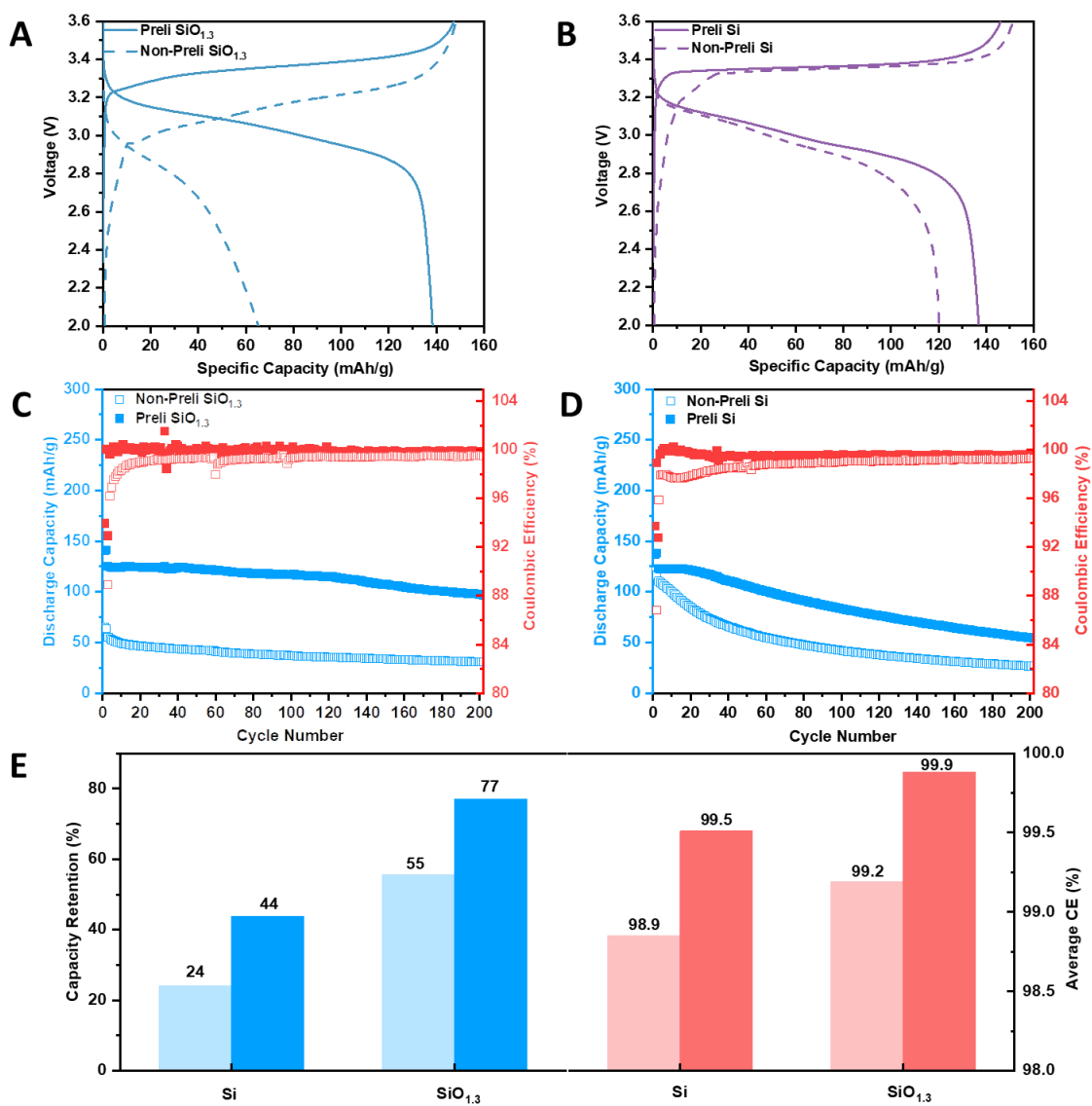

**Figure 3.** (A) Initial cycle voltage profile and (C) long-term cycle performance of prelithiated and non-prelithiated $SiO_{1.3}$ in full cell with LFP cathode. (B) Initial cycle voltage profile and (D) long-term cycle performance of prelithiated and non-

prelithiated μSi in full cell with LFP cathode. (E) Capacity retention and average CE comparison of prelithiated and non-prelithiated $SiO_{1.3}$ and μSi in full cell.

A key challenge for the application of Si is the volume expansion (~300%) during its lithiation, which leads to the disruption of the SEI layer, continuous electrolyte consumption due to newly formed SEI, and electrode pulverization.[32,33] Cross section images were collected using scanning electron microscopy (FIB-SEM) to evaluate electrode thickness after the cycling in full cells, as shown in **Figure 4**. After 100 cycles, the prelithiated μSi anode thickness increases relatively by 28% compared with that after the initial cycle (12.4 μm versus 9.7 μm). It should be noted this volume expansion is not trivial considering the thickness evaluation is performed at the delithiation state of the anode. In contrast, the thickness of the prelithiated $SiO_{1.3}$ anode only increases relatively by 3% after 100 cycles (**Figure 4A and 4B**). The much lower volume expansion of $SiO_{1.3}$ can be ascribed to both the bulk sub-oxide matrix and interface stabilization. Electrochemical impedance spectroscopy (EIS) of the full cell was then conducted to explore the interface impedance changes during extended cycles. **Figure 4C and 4F** show the Nyquist plots of the full cells prepared using the prelithiated anodes after the initial cycle and after 100 cycles, as well as the corresponding fitting plots using the equivalent circuit. R1, R2, and R3 represent electrolyte resistance, SEI resistance, and interface charge transfer resistance, respectively.[34] It turns out the interfacial resistance (R2 and R3) increases dramatically from 18 Ω to 135 Ω after 100 cycles (**Figure S6**) for the full cells using prelithiated μSi anodes. While a slight increase (29 Ω) is observed for the prelithiated $SiO_{1.3}$ anodes after 100 cycles. Since the LFP cathode side is relatively stable in the full cell, the EIS analysis well manifests the interface of prelithiated $SiO_{1.3}$ is more intact than the μSi over long-term cycling.

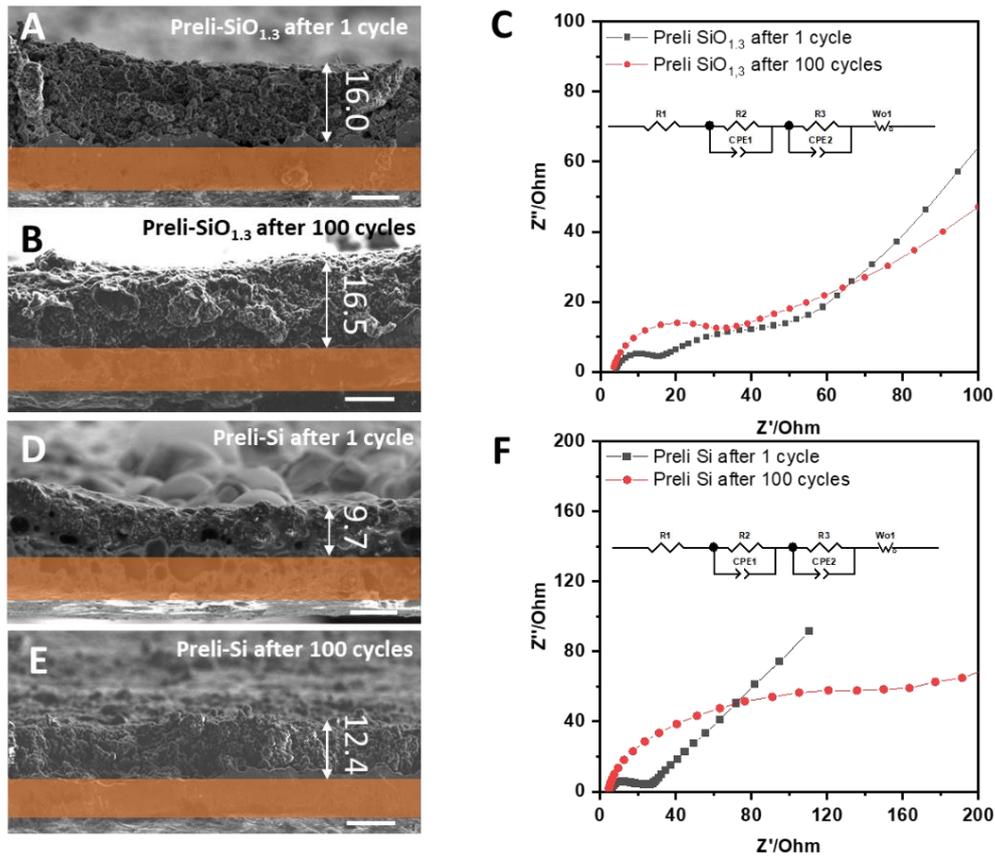

**Figure 4.** SEM cross section images of prelithiated SiO$_{1.3}$ after (A) 1 cycle and (B) 100 cycles and (C) the impedance change of full cells upon cycling. SEM cross section images of prelithiated μSi after (D) 1 cycle and (E) 100 cycles and (F) the impedance change of full cells upon cycling. All the scale bars in SEM images represent 10 μm. The orange bar in each SEM image indicates the location of the Cu current collector.

**SEI Composition and Structure of Prelithiated Si-based Anodes**

Further analysis of SEI composition and structure at the nanoscale is then performed to understand the prelithiation effect on the interface stabilization of SiO$_{1.3}$ anode. In order to compare the SEI formed in the electrochemical reaction, the pristine SiO$_{1.3}$ anode discharged in the half cell to the same potential as the prelithiated sample was also prepared. To prevent the washing effect on removing the fragile and/or reactive SEI components such as carbonate and oxides,[35] no solvent was introduced in all the samples preparation process for (S)TEM and energy dispersive X-ray (EDX) analysis.

As shown in **Figure 5A and 5B**, the SEI formed through the prelithiation process shows a mosaic microstructure with uniform elemental distribution including C, F, Si, and P. In contrast, a layered microstructure is observed for the SEI formed in the electrochemical process, where C, F, and P are in the outer layer of the SEI and the inner layer mainly consists of Si compounds. This obvious difference can be confirmed and quantified by the EDX line scan analysis, as shown in **Figure S7**. EDX Si line scan results in **Figure S7** clearly indicate the formation of different microstructures of SEI through distinct lithiation processes. In the typical electrochemical process, slow C-rate is applied for the lithiation process so that different SEI components will form close to the equilibrium conditions. As predicted by the differential capacity curves in **Figure 1C**, the electrolyte decomposition byproducts, such as $Li_2CO_3$ and LiF, appear on the surface of particles first. And then lithium ions will pass through this outer layer to trigger the conversion of $SiO_2$ to lithium silicates and oxides to form the inner layer of SEI. While the short-circuit prelithiation process drives concentrated $Li^+$ flux together with strong electric field around the contact points, which results in the SEI formation far away from the thermodynamically stable state. All the SEI reactions almost occur simultaneously to construct the observed mosaic microstructure.

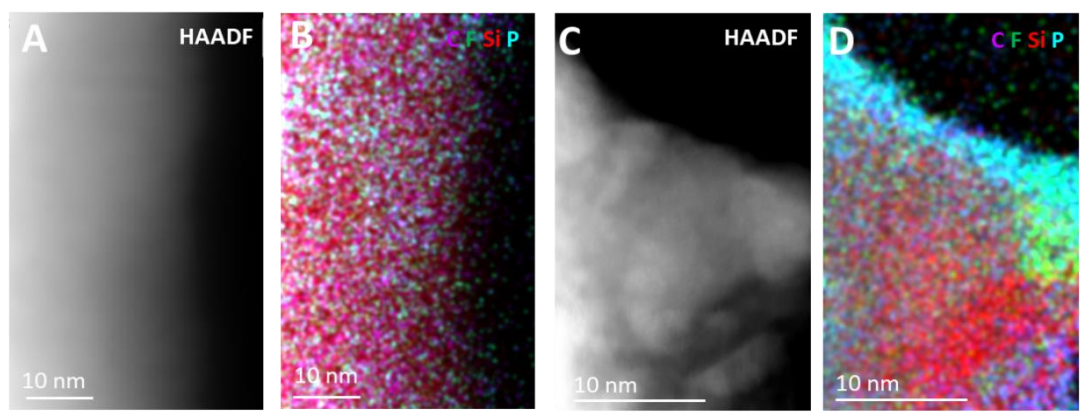

**Figure 5.** (A) HAADF image of prelithiated $SiO_{1.3}$ using the short-circuit electrochemical method and (B) its corresponding EDX mapping. (C) HAADF image of lithiated $SiO_{1.3}$ in the half cell to the same potential as the prelithiated sample and (D) its corresponding EDX mapping.

High resolution transmission electron microscopy (HRTEM) was then performed to detect the crystalline component of the SEI for both prelithiated $SiO_{1.3}$ and μSi. To protect thin surface layer from electron dose damage, all the HRTEM images in this work were recorded under cryogenic temperature, following our best sample transfer and imaging protocols.[36] The HRTEM image and the corresponding fast Fourier transform (FFT) pattern of the pristine and prelithiated $SiO_{1.3}$ electrodes are shown in **Figure 6A and 6B**. In the pristine state of $SiO_{1.3}$, the amorphous phase is found to be dominant with nanosized Si (as highlighted with the white area in the HRTEM image) based on the FFT spots of (111) plane of the diamond cubic structure. The pristine $SiO_{1.3}$ particle can thus be described as a mixture of nanosized Si and amorphous $SiO_2$, which is consistent with our XRD analysis. After the prelithiation, the inset FFT pattern in **Figure 6B** illustrates the coexistence of LiF, $Li_2CO_3$, $Li_2O$, $Li_4SiO_4$, $Li_2SiO_3$, and Li-Si alloy species distributed in the interface area by matching the lattice spacings (**Table S2**) of corresponding species with the pattern. All the identified prelithiation products correspond well with the differential capacity analysis for the three distinctive reactions. The uniform mixing of these products in the interface region supports the mosaic model proposed based on the EDX analysis. While the SEI formed in the typical electrochemical process with slow C-rate contains electrolyte decomposition byproducts (LiF, $Li_2CO_3$, etc.) as the outer layer and lithium silicate ($Li_4SiO_4$) as the inner layer, as shown in **Figure S8**. This observation is consistent with the layered microstructure of SEI under the equilibrium conditions. It is then as expected, the prelithiated μSi also illustrates the mosaic microstructure of SEI due to the direct contact (**Figure 6C and 6D**). Compared with the prelithiated $SiO_{1.3}$, only $Li_4SiO_4$ is identified in the interface of prelithiated μSi. Note the lithium silicate in the prelithiated μSi is originated from the intrinsic thin oxide layer on the pristine sample (**Figure 6C**).

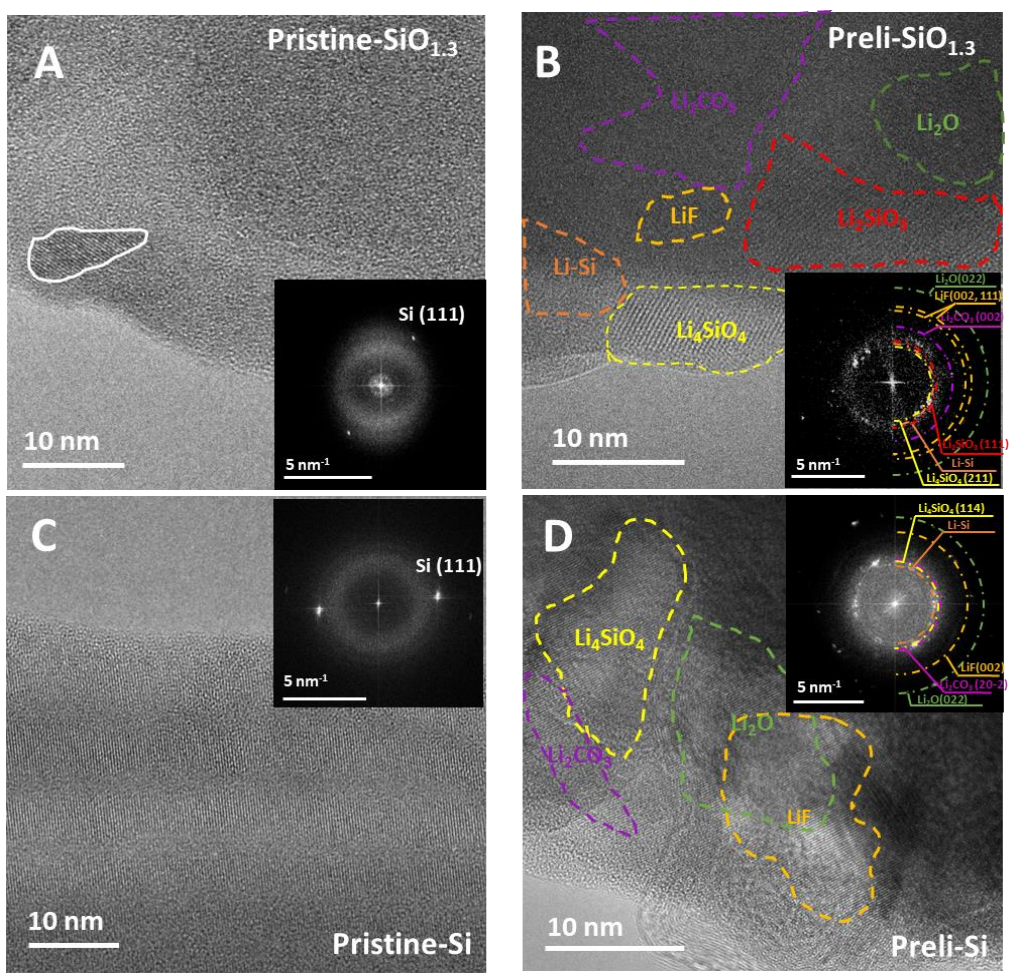

**Figure 6.** HRTEM images of pristine (A) and prelithiated (B) SiO$_{1.3}$ with corresponding FFT patterns. HRTEM images of pristine (C) and prelithiated (D) μSi with corresponding FFT patterns.

We then performed XPS depth profiling on the surface of the SiO$_{1.3}$ and μSi anode after the prelithiation to analyze the chemical composition of SEI layers. The results in two core levels of Si 2p and O 1s are shown in **Figure 7**. Both samples were profiled using high energy large Ar$^+$ clusters (10 keV) so that the depth profile was completed with a sputter rate of 2 nm/min. The binding energy values of various Li$_x$SiO$_y$ phases were determined in previous studies of Li$_2$SiO$_3$ and Li$_4$SiO$_4$.[37] It is important to note that a wider range of binding energy was used to fit Li$_2$SiO$_3$ and Li$_4$SiO$_4$ peaks as an approximation to interpret bonding distribution in the amorphous matrix. After the prelithiation, the Si 2p spectra in **Figure 7** show that the interface of SiO$_{1.3}$ is primarily composed of lithium silicates (101.3 eV for Li$_4$SiO$_4$ and 102.8 eV for Li$_2$SiO$_3$). A much higher Li-Si alloy peak is observed in the prelithiated μSi, indicating the formation of a less conformal SEI layer. Only one type of lithium silicate (Li$_4$SiO$_4$) is identified in the SEI of prelithiated μSi, which is consistent with the HRTEM results. For O 1s spectra, the peaks at 532.0 eV and 529.0 eV are assigned to Li-Si-O and Li$_2$O species, respectively.[38] A larger amount of Li$_2$O can be found in the prelithiated SiO$_{1.3}$ than μSi due to the higher oxygen content in the structure. Other common components of SEI (LiF, P-F-O, Li$_2$CO$_3$) can also be identified in F 1s and C 1s spectra, as shown in **Figure S9**. No obvious difference of these electrolyte decomposition byproducts is found in the SEI of prelithiated SiO$_{1.3}$ and μSi.

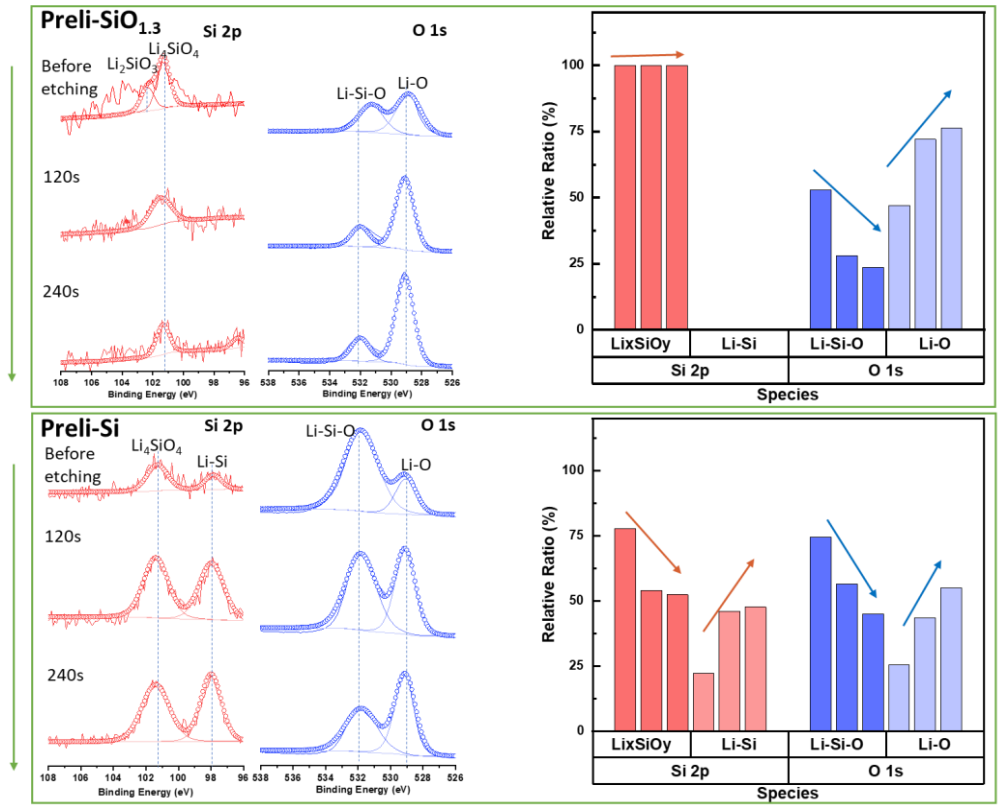

**Figure 7.** The XPS depth profiling of Si 2p and O 1s spectra and quantitative analysis of different SEI components relative ratio for the prelithiated SiO$_{1.3}$ and μSi anodes.

**Discussion on Mechanism of Cycling Performance Improvement via Prelithiation**

The above (S)TEM and EDX results reveal different SEI formation pathways for prelithiation process by direct contact compared with the conventional electrochemical process. After prelithiation, a mosaic type SEI containing lithium silicates is formed for both $SiO_{1.3}$ and μSi anodes. While the electrolyte decomposition byproducts (LiF, $Li_2CO_3$, etc.) are dominant species on the surface of the anode particles through the slow C-rate discharge to the same lithiation potential. The ideal SEI requires good electronic resistivity and ionic conductivity, which is critical for interface stabilization over long-term cycling. Therefore, the electronic and ionic conductivity of several identified lithium silicates and electrolyte decomposition byproducts were measured and the results are shown in **Table 1**. Lithium silicates exhibit at least two orders of magnitude better Li ion conductivity than that of LiF or $Li_2CO_3$. And the relatively high ionic-to-electronic conductivity ratio (~$10^3$) of lithium silicates makes it sufficient to inhibit the unwanted electrochemical reactions. More importantly, compared with LiF and $Li_2CO_3$, the much higher mechanical strength of lithium silicates enables robust surface layer to sustain drastic volume fluctuation.

**Table 1.** Major SEI components identified in prelithiated $SiO_{1.3}$ and μSi together with the measured physical properties of each component.

| SEI components | Electronic conductivity (S/cm) | Ionic conductivity (S/cm) | Ionic-to-electronic conductivity ratio | Young's modulus (GPa) | $SiO_{1.3}$ | Si |
|---|---|---|---|---|---|---|
| LiF | ~$10^{-12}$ [39] | ~$10^{-8}$ [39] | ~$10^4$ | 58.1-125 [40] | √ | √ |
| $Li_2CO_3$ | ~$10^{-10}$ [41] | ~$10^{-9}$ [42] | ~10 | 36.2-54.8 [40] | √ | √ |
| $Li_2O$ | ~$10^{-14}$ [43] | ~$10^{-12}$ [43] | ~$10^2$ | 163.58 [44] | √ | √ |
| $Li_4SiO_4$ | $1.66\times10^{-9}$ | $8.57\times10^{-7}$ | ~$10^3$ | 108.2-125.8 [45] | √ | √ |
| $Li_2SiO_3$ | $1.72\times10^{-9}$ | $3.566\times10^{-6}$ | ~$10^3$ | 110.7-126.7 [45] | √ | × |
| $Li_2Si_2O_5$ | $5.61\times10^{-10}$ | $2.778\times10^{-7}$ | ~$10^3$ | 100.3 [46] | √ | × |

Compared with the prelithiated μSi, higher ratio of lithium silicates and lithium oxide exist in the SEI of the prelithiated $SiO_{1.3}$, which further improves the interface stability. As demonstrated in **Figure S10**, the mosaic SEI microstructure and silicates enriched composition well maintain even after 180 cycles. The appearance of $Li_2Si_2O_5$ compound in the SEI after 180 cycles is consistent with the previous study on the structural evolution of SiO$_x$ thin-film electrodes with subsequent lithiation/delithiation cycles.[29] Additionally, no bulk particle cracking is observed for the prelithiated SiO$_{1.3}$ after 100 cycles (**Figure S11**). As summarized in the following schematics (**Figure 8**), the mosaic microstructure enabled by prelithiation is well maintained after long cycling in SiO$_{1.3}$. While for the μSi, the particle cracking is inevitable due to the large volume expansion (~ 300%) and limited amount of lithium silicates in SEI, which will result in the cycling failure eventually. Therefore, the preferred SEI microstructure and composition together with the lower volume expansion (< 150%) result in the superior cycling performance of prelithiated SiO$_{1.3}$ over μSi.

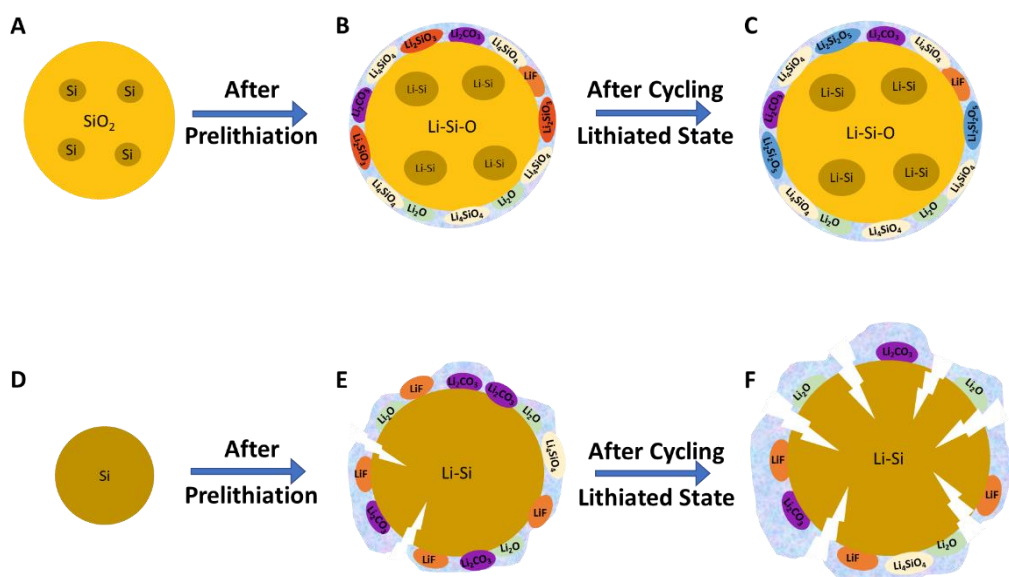

**Figure 8.** Schematics of prelithiation effect on cycling stability for both SiO$_{1.3}$ and μSi.

## CONCLUSION

In summary, through a combined imaging, spectroscopy, and electrochemical analysis approach, this study reveals a mixed mosaic microstructure interface with better electrochemical and mechanical properties formed during the prelithiation process on the Si-based anodes using the direct contact method. This prelithiated interface is composed of lithium silicates, lithium oxide, and other typical electrolyte decomposition byproducts. All of these components are uniformly mixed in the interface to form a mosaic microstructure. Due to the high ionic-to-electronic conductivity ratio and mechanical strength, lithium silicates enriched SEI is expected to enable long-term cycling stability. With more oxygen content to participate in the SEI formation process, the prelithiated $SiO_{1.3}$ anode based full cells with LFP cathode exhibit an initial specific discharge capacity of 138 mAh/g and an ICE of 94% at the current density of C/10. Moreover, 77% capacity retention is obtained after 200 cycles of charging and discharging at the current density of C/3. The ICE, specific capacity, and cycling stability of lithiated $SiO_{1.3}$-based full cells are improved significantly, indicating great prospects for the commercial application of Si-based anode materials.

# EXPERIMENTAL PROCEDURES

## Sample Preparation

The pristine anode on a copper foil is composed of 70 *wt.* % silicon monoxide (99.9%, Alfa Aesar, active material) or micron silicon (1-5 μm, Alfa Aesar), 20 *wt.* % acetylene black (AB, Denka, conductive carbon), and 10 *wt.* % Poly (acrylic acid) (PAA, Mv 450,000, Sigma Aldrich, binder). The anode laminates were punched into 13 mm diameter discs with mass loading of ~1.2 mg/cm$^2$ for SiO$_{1.3}$ and ~0.8 mg/cm$^2$ for Si. The LiFePO$_4$ (LFP) cathode for the full cell was purchased from NEI (Areal capacity ~ 1.25 mAh/cm$^2$, diameter 1/2 inch). 1 mol/L LiPF$_6$ was dissolved in ethylene carbonate (EC): dimethyl carbonate (DEC) (1:1, v/v) with 10% of fluoroethylene carbonate (FEC) (Gotion, USA) as the electrolyte.

## Prelithiation Setup

Anode was wetted with 30 μL electrolyte and directly contact with Li metal chip (1 mm thick, 16 mm in diameter). 75 g weight was added to the top of the setup to ensure close contact between the SiO$_{1.3}$ electrode and Li metal chip during the prelithiation process with time monitor. Various prelithiated electrodes were assembled to the CR-2032 coin cell with Li metal as counter electrode and performed electrochemical testing. ICE ~100% is considered as fully prelithiated anode condition.

## Electrochemical Measurements

The fully-prelithiated anodes and LFP cathodes (N/P ratio 1.1-1.2) with 50 μL electrolyte were assembled in an Ar-filled glove box (H$_2$O < 0.1 ppm). The galvanostatic charge/discharge test was carried out at C/10-rate (1 C =170 mAh/g) for the initial two cycles in the voltage range of 2.0-3.6 V. The cells were then charged and discharged at C/3-rate for the rest of the cycles. All the tests were performed at room temperature. After cycling, the cells were dissembled in the Ar-filled glove box and the anode were rinsed with DMC solvent to remove salt residue on the surface. After drying at room temperature, the cycled anode materials were stored in the glove box for further characterizations.

**Conductivity Test**

The sample powders were pressed into a 10 mm-diameter pellets at 625 MPa with the thickness measured. The EIS and DC polarization measurements were conducted at room temperature on Biologic using two titanium blocking electrodes enclosed by a polyetheretherketone holder. An applied AC potential of 30 mV over a frequency range from 1 MHz to 1 mHz was used for the EIS measurement. A constant voltage of 0.1 V was applied for DC polarization. The stabilized current can be obtained when the ionic transportation is fully eliminated.

**Scanning Electron Microscope (SEM)**

The experiment was carried out at 10 keV as the operating voltage at 0.1 nA with ETD detector under standard mode. The sample was prepared in the Ar-filled glovebox and transferred to the SEM chamber with the minimum exposure to air for the thickness check.

**X-ray Diffraction (XRD)**

The XRD with Cu target was carried out using the capillary to eliminate the reaction from the air. The scanning region is 10-90 degree at 4 degree/min.

**Transmission Electron Microscope (TEM)**

The samples were carefully scratched from the electrode and dispersed to the TEM grid (carbon side). A single-tilt liquid nitrogen cooling holder (Gatan 626) was used to cool the samples to approximately -170°C to minimize electron beam damage where the TEM grids were sealed in heat-seal bags and transferred to TEM column using a purging home-made glovebox filled with Ar gas. TEM images, and selected area electron diffraction (SAED) patterns were conducted on a JEOL JEM-2800F TEM, equipped with a Gatan Oneview camera operated at 200 kV. Pristine $SiO_{1.3}$ images were obtained on ThermoFisher Talos X200 equipped with a Gatan Oneview camera operated at 200 kV and UltraFast DualEELS Spectrum Imaging detector. The image was acquired with minimum beam damage at spot size 6 with a dose rate of 200 electrons/Å$^2$/s. The EELS spectrum and mapping were collected with an exposure time of 0.02 s, and the dispersion energy was 0.25 eV per channel.

**X-ray Photoelectron Spectroscopy (XPS)**

XPS was performed in an AXIS Supra XPS by Kratos Analytical. XPS spectra were collected using a monochromatized Al Ka radiation (hy = 1,486.7 eV) under a base pressure of $10^{-9}$ Torr. To avoid moisture and air exposure, a nitrogen filled glovebox was directly connected to XPS spectrometer. Survey scans were performed with a step size of 1.0 eV, followed by a high-resolution scan with 0.1 eV resolution, for Li 1s, C 1s, O 1s, F 1s and Si 2p regions. All spectra were calibrated with adventitious carbon 1s (284.6 eV) and analyzed by CasaXPS software. The etching condition used was $Ar^+$ mono mode, 10 keV voltage. The etching intervals was 120 s.

## AUTHOR CONTRIBUTIONS

## ACKNOWLEDGEMENTS

The authors gratefully acknowledge funding supported by the Advano University Research Program and Materials Science and Engineering program of UCSD. SEM was performed at the San Diego Nanotechnology Infrastructure (SDNI), a member of the National Nanotechnology Coordinated Infrastructure, which is supported by the National Science Foundation (grant ECCS-1542148). The authors acknowledged the UC Irvine Materials Research Institute (IMRI) for the use of the XPS and STEM-EDX, funded in part by the National Science Foundation Major Research Instrumentation Program under Grant CHE-1338173. The authors would like to acknowledge Neware Technology Limited for the donation of BTS4000 cyclers which are used for testing the cells in this work. Special thanks to Dr. Ich Tran for the assistance on the XPS data acquisition and discussion.

## DECLARATION OF INTERESTS

The authors declare no competing interests.